\documentclass[12pt]{article}
\usepackage{amssymb,amsfonts}
\usepackage{latexsym}
\usepackage{amsmath, amssymb}
\usepackage{amsfonts}
\usepackage{dsfont}
\usepackage{marvosym}

\usepackage[usenames,dvipsnames,svgnames,table]{xcolor}

\usepackage{slashed}
\usepackage{graphicx}
\usepackage{hyperref}
\usepackage{array}   
\usepackage{ulem}
\usepackage{epsfig}
\usepackage{psfrag}

\usepackage{mathtools,cancel}

\usepackage{color}
\usepackage{cite}
\usepackage{geometry}
\makeatletter\@addtoreset{equation}{section}\makeatother

\newcommand{\beq}{\begin{equation}}
\newcommand{\eeq}{\end{equation}}
\newcommand{\bal}{\begin{equation}\begin{aligned}}
\newcommand{\eal}{\end{aligned}\end{equation}}

\newcommand{\CGR}{\textcolor{Grey}}

\newcommand{\ol}{\overline }
\newcommand{\dd}{\text{d} }

\newcommand{\he}{\hat{e}}
\newcommand{\CP}{\mathbb{C}\text{P} }
\newcommand{\ts}{\tilde{\sigma}}
\newcommand{\Lie}{\text{Lie}}

\newcommand{\mf}{\mathfrak}
\newcommand{\mbb}{\mathbb}
\newcommand{\mc}{\mathcal}
\newcommand{\ann}{\text{ann}}
\newcommand{\Ad}{\text{Ad}}

\newcommand{\address}[1]{\vbox{\center\em#1}}
\renewcommand{\title}[1]{\vbox{\center\huge{#1}}\vspace{5mm}}

\graphicspath{{figures/}}

\begin{document}

\newcommand{\penguin}{\includegraphics[width=0.2in]{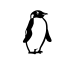}}
\makeatletter
\def\@fnsymbol#1{\ensuremath{\ifcase#1\or \,\or \text{\Cat}  \or
   \mathsection\or \mathparagraph\or \|\or **\or \dagger\dagger
   \or \ddagger\ddagger \else\@ctrerr\fi}}
\makeatother
\renewcommand{\thefootnote}{\fnsymbol{footnote}}
\setcounter{footnote}{0}

\begin{center}
\phantom{xx}

\vspace{20mm}

\title{Adjoint orbits, generalised parallelisable spaces and consistent truncations}

\vspace{10mm}

     Louise Anderson\penguin  \footnote{ \hspace{-4mm}\protect \penguin \hspace{-1mm}\href{mailto:louise.m.a.anderson@imperial.ac.uk}
 { louise.m.a.anderson@imperial.ac.uk}}~
 \\
\address{The Blackett Laboratory, Imperial College London,\\
Prince Consort Road, London SW7 2AZ, United Kingdom}

\end{center}

\vspace{8mm}
\abstract{
\normalsize{
\noindent
The aim of this note is to present some new explicit examples of $O(d,d)$-generalised Leibniz parallelisable spaces arising as the normal bundles of adjoint orbits $\mc{O}$  of some semi-simple Lie group $G$. 
Using this construction, an explicit expression for a generalised frame is given in the case when the orbits are regular, but subtleties arise when  they become degenerate. In the case of regular orbits, the resulting space is a globally flat fiber bundle over $\mc{O}$ which can be made compact, allowing for a generalised Scherk-Schwartz reduction. This means these spaces should admit  consistent supergravity truncations.  For degenerate orbits, the procedure hinges on the existence of a suitable metric, allowing for a consistent normalisation of the generalised frame.

}}

\renewcommand{\thefootnote}{\arabic{footnote}}
\setcounter{footnote}{0}

\section{Introduction}
The story of consistent truncations is a long one within supergravity, and classifying all possible consistent truncations remains an open problem. The classical examples of Scherk-Schwartz reductions \cite{Scherk:1978ta,Scherk:1979zr} on local group manifolds \cite{Duff:1986, Cvetic:2003jy}, together with the exceptional consistent truncations on spheres \cite{deWit:1986oxb,Nastase:1999cb,Nastase:1999kf,Cvetic:2000nc,Cvetic:2000ah} were unified as \emph{generalised} Scherk-Schwartz reductions\footnote{For other examples of the use of generalised geometry and exceptional field theory in the context of generalised Scherk-Schwartz reductions, see \cite{Berman:2012uy,Musaev:2013rq,Aldazabal:2013mya,Baguet:2015sma,deWit:2013ija, Godazgar:2013oba, Godazgar:2013nma, Godazgar:2013dma, Godazgar:2013pfa}. } in \cite{Lee:2014mla} using the language of generalised geometry. 

The well-known result of local group manifolds critically hinges on them being (Liebniz) parallelisable. Using  generalised geometry, this requirement can be relaxed to  \emph{generalised} Leibniz parallelisability   of the space  \cite{Grana:2008yw}, which allows for consistent truncations on  certain homogenous spaces, such as spheres\cite{Lee:2014mla}, twisted tori and hyperbeloids\cite{Hohm:2014qga, Malek:2015hma}\footnote{And certain products thereof \cite{Baguet:2015iou, Inverso:2016eet, Malek:2017cle}. Preserving fewer supersymmetries, more general consistent truncations are possible, see for example \cite{Malek:2016bpu, Malek:2017njj}. }. These results makes the question of which manifolds admit a generalised Leibniz parallelisation an interesting one. In \cite{duBosque:2017dfc}, a first step towards a systematic construction of these spaces, and the generalised frames on them, was taken using exceptional field theory\cite{Berman:2010is, Berman:2011jh, Berman:2012vc, Hohm:2013vpa, Hohm:2013uia, Hohm:2014fxa, Abzalov:2015ega, Musaev:2015ces}. The aim of this note is to further contribute to this by presenting some examples of generalised parallelisable spaces and explicit constructions of suitable generalised frames on them  in the case of $O(d,d)$ generalised geometry \cite{Hitchin:2004ut,Gualtieri:2004thesis}.

By considering spaces that can be thought of as adjoint orbits, $\mc{O}(a)$, of some semi-simple Lie group $G$, with corresponding stabiliser groups $H_a$, we show the normal bundles over these orbits, $ N\mc{O}(a)$, in the case when $\mc{O}(a)$ is regular, are generalised parallelisable  in the sense of  \cite{Grana:2008yw}.
An explicit $O(d,d)$-generalised frame is given for the generalised tangent bundle of this space, and a two-form $B$ is constructed such that this frame satisfies the Leibniz conditions.
 In the case where the orbit is degenerate, which is equivalent to the stabiliser being non-abelian, construction of such a  frame critically hinges on the existence of a suitable metric (see section \ref{sec:metric}). 
When the orbits are regular, the normal bundle is a flat bundle over the orbit and can be made compact,  and they  should thus admit a generalised Scherk-Schwartz reduction of type IIA, IIB and bosonic string theory 
\cite{Coimbra:2011nw}\footnote{For the closely related double field theory formalism, see for example \cite{Siegel:1993xq, Siegel:1993th, Hohm:2010xe, Jeon:2010rw, Jeon:2011cn, Hull:2009mi, Hull:2009zb, Hohm:2010jy, Hohm:2010pp, Hassler:2016srl}}.

 Recently, an exposition of generalised Scherk-Schwartz uplifts of gauged supergravities was presented in \cite{Inverso:2017lrz}, and generic classes of generalised Leibniz parallelisable spaces were presented. The spaces obtained here from regular adjoint orbits should fall under this classification.

The structure of this note is as follows: In the next section, the setup and conventions used herein will be presented, after which  an action of $G\ltimes \mathfrak{g}$ on the normal bundle of the adjoint orbits will be constructed.  In section 3, this will be used to explicitly construct a generalised Leibniz parallelisation of the $O(d,d)$-generalised tangent bundle of $ N\mc{O}(a)$. In section 4, some explicit examples are presented, and section 5 concludes with a discussion.

\section{Adjoint orbits and a $G\ltimes \mf{g}$ action}
 (Co-)adjoint orbits have been extensively studied in the literature (see for example chapter 8 of \cite{Besse:1987 } or for a recent review, see  \cite{Crooks:2017}), and viewing the homogenous spaces $G/H_a$ as such orbits allows us to take advantage of known results about their properties, and especially their normal bundles. The adjoint orbits of a group $G$ have at most dimension $\dim(G)-\text{rank}(G)$, and is then known as regular orbits. Orbits with smaller dimensions are said to be degenerate. Regular orbits are isoparametric submanifolds of $G$, which immediately gives that the normal bundles  to these orbits are flat \cite{terng1985}.


\subsection{Conventions and preliminaries}
Let $G$ be a semi-simple Lie group, and $H$ a subgroup of $G$, with corresponding Lie algebras $\mf{g}, \, \mf{h}$. We can as usual identify $\mf{g}$ with the tangent space at the origin of the group $G$. Furthermore, there is an  $Ad_G$-invariant scalar product on $\mf{g}$, denoted by $\langle \cdot, \cdot \rangle$.

Let $a$ be an element of the Lie algebra $\mf{g}$. The orbit of $a$ under the adjoint action of $G$ is denoted $\mc{O}(a)$. Each orbit arising in this way may be parametrised by an element in the (closure of) the fundamental Weyl chamber of $G$, denoted by $C\; \left(\ol{C}\right)$.  Without loss of generality, we will always take $a\in \ol{C}$.  The regular orbits are given by $a\in C$, whereas for $a\in \ol{C} \setminus C$, the orbit is degenerate. The different classes of orbits are therefore completely determined by the fundamental Weyl chamber of the group.

Consider the adjoint orbit through a point $a\in \ol{C}\subset \mf{g}$, and let $H$ be the stabiliser subgroup of $a$. The orbit $\mc{O}(a)$ is then isomorphic to the homogenous space $G/H$, and the Lie algebra $\mf{h}$ can be identified with the annihilator of $a$, i.e. 
\beq
T_e H \; = \; \ann (a) \;= \; \left\{ \xi \in \mf{g} , \; [\xi, a] = 0 \; \right\} \; \cong  \; \mf{h}
.\eeq 
There is a natural embedding of  $\mc{O}(a)$ in $\mf{g}$ .   The $\Ad_G$-invariant inner product on $\mf{g}$ restricted to the orbit induces an $Ad_G$-invariant inner product on  $\mc{O}(a)\hookrightarrow \mf{g}$,  allowing us to identify the tangent- and the cotangent spaces, i.e. $T^*(G/H)  \simeq T(G/H)$. 

With respect to this inner product, there is an orthogonal decomposition of $\mf{g}$ as:
\begin{align}
\label{eq:orth_decomp}
\mf{g} = \mf{h}\oplus \mf{v}
,\end{align}
which means that we can identify $T_{\Pi(e)} \mc{O}(a)$ with $\mf{v}$, where $\Pi: G\longrightarrow G/H$ is the canonical projection.

We will here be interested in the normal bundle over this orbit, $N\mc{O}(a)$, which we view as $N\mc{O}(a) \hookrightarrow \mf{g}\otimes \mf{g}$. That is, we parametrise the space  by pairs $(x,\,n)\in \mf{g}\otimes \mf{g}$, such that $x$  parametrise the orbit and $n$ the normal directions.

It is clear that we at any point $x=\Ad_g a$ in this orbit will have: 
\begin{align*}
T_x\mc{O}(a) \, \simeq& \;\, \mf{v}
\qquad,\qquad
N_x\mc{O}(a) \, \simeq  \, \mf{h}
,\end{align*}
by equation \eqref{eq:orth_decomp}.

 In the rest of this section, we will show that we can construct an action of $G\ltimes \mf{g}$ on the normal bundle. In the next section, we will show that, under certain assumptions, this is enough for constructing a generalised frame on this space, allowing for a generalised Scherk-Schwartz reduction. These assumptions are trivially fulfilled in the case when $\mc{O}(a)$ is regular, in which case we present an explicit construction of the generalised frame. However, for degenerate orbits, the situation is more subtle and the existence of a generalised frame appears less clear.


\subsection{An action of $G\ltimes \mf{g}$ on $\mc{M}$}

Consider now a point in the normal bundle $N\mc{O} \hookrightarrow \mf{g}\otimes \mf{g}$, given by:
\begin{align*}
(x,n) &\qquad x\in \mf{g}, \;\; n\in \ann(x)\simeq\mf{h}
.\end{align*}
(We will from now on omit the $a$ and only denote the orbit only by  $\mc{O}$ for brevity.) In \cite{mathph-0609005}, the authors showed that there is a natural action of $G\ltimes \mf{g}$ on the tangent space, $T_x\mc{O}$. This can be extended to the  normal bundle as follows: let  $(g,\eta) \; \in \; G\ltimes \mf{g}$, where the $\mf{g}$ should be viewed as a vector space, and the product be given by:
\beq
(g',\eta' )(g, \eta) \;= \; (g'g, \Ad_{g'}\eta+\eta')
.\eeq 
The bracket in the Lie algebra of $G\ltimes \mf{g}$ is given by:
\begin{align}
\label{eq:Lie_bracket}
\left[ \, (\gamma_a,\tilde{\gamma}_a) \, ,\, (\gamma_b,\tilde{\gamma}_b) \right]
\,= \,
\left( \,
[\gamma_a,\gamma_b]\, 
, \,
[\gamma_a, \tilde{\gamma}_b]- [\gamma_b,\tilde{\gamma}_a]
\,\right)
,\end{align}
where $(\gamma_a,\tilde{\gamma}_a)\in \Lie\left(G\ltimes \mf{g}\right)$.

We then define the action on $N_x\mc{O}$ by:
\begin{align}
\label{eq:action}
(g,\eta)(x,n) \; =& \; \big(\Ad_g x, \;  \Ad_g n +   \Pi_{\ann(\Ad_g x)}   \eta  \big) 
,\end{align}
where $\, \Pi$ denotes the orthogonal projection with respect to the $\Ad_G$-invariant inner product $\langle \cdot, \cdot \rangle$.


\subsubsection*{The infinitesimal action}
Let $\sigma_a, \ts_a$ be a basis for the Lie algebra $\Lie\left(G\ltimes \mf{g}\right)$. The action of \eqref{eq:action} give rise to $2 \dim G$ vector fields, which in terms of the basis elements $\sigma_a, \ts_a$ can be written as:
\beq
\he_{a,b} \, = \,  \big(  [\sigma_a, x]   , \;  [\sigma_a, n ]  +   \, \Pi_{\ann(x)}  \ts_b  \; \big) 
.\eeq
Let the vector fields generated by the $\sigma$'s be denoted by $\he^v_a$ and the ones generated by the $\ts$'s by $\he^h_a$.  In the case where  $\ann(x)$ is abelian, (i.e. for a regular orbit $\mc{O}$), these will be completely orthogonal in the sense of \eqref{eq:orth_decomp}. In this case, we have  $\he_a^{v} \in \ann(x)^\perp\simeq \mf{v}$ and $\he_a^{h}\in \ann(x)\simeq \mf{h}$ respectively, hence the notation.
These satisfies the commutation relations of equation \eqref{eq:Lie_bracket}:
 \begin{align}
\label{eq:Lie_rel}
[\he^v_a, \he^v_b] \; = \; f_{ab}\,^c \he^v_c
\qquad,\qquad
[\he^v_a, \he^h_b] \; = \; f_{ab}\,^c \he^h_c
\qquad,\qquad
[\he^h_a, \he^h_b] \; = \; 0
.\end{align}
However, even though their commutation relations are straight-forward, these vectors are not globally defined and cannot be frames for any viable metric. We can however, using their sums and differences,  create  two sets of $\dim G$ vectors that are globally defined, denoted by $\he_a^{\pm}$:
\beq
\label{eq:vecs}
\he_a^{\pm} \, = \,\he^v_a \pm \he^h_a \, = \,  \big(  [\sigma_a, x]   , \;  [\sigma_a, n ] \pm   \, \Pi_{\ann(x)}  \sigma_a  \; \big) 
.\eeq
It is easy to see that these indeed are globally defined, since $[\sigma_a,x]$ will only vanish when $\sigma_a\in \ann(x)$, meaning that $\, \Pi_{\ann(x)}\sigma_a = \sigma_a \neq 0$.
They satisfies the commutation relations:
 {\small
 \begin{align}
 \label{eq:comm_rels_pm}
 [\he^{\pm}_a, \he^{\pm}_b]  =& 
\frac{1}{2}  f_{ab}\,^c \left(
    3  \he^\pm_c 
-     \he^\mp_c 
\right)
\qquad,\qquad
 [\he^+_a, \he^-_b]  =
\frac{1}{2} f_{ab}\,^c \left(
 \he^+_c  +  \he^-_c  
 \right)
 \end{align}
}

 \vspace{-0.7cm}
 \subsection{The metric \label{sec:metric}}
 We can define a metric using the vectors $\he^\pm$ by requiering:
 \beq
 \label{eq:mertic}
 g(\he^\pm_a, \he^\pm_b) =\delta_{ab}
 .\eeq
 For this to hold for both sets of vectors, the metric must satisfy 
  \begin{equation}
  \label{eq:cross_terms}
 g(\he^v_a, \he^h_b)+ g(\he^h_a, \he^v_b) =0
 .\end{equation}
 Consider $N\mc{O}\hookrightarrow \mf{g}\otimes \mf{g}$, and assume the metric is block diagonal in the two $\mf{g}$-subspaces. 
If the orbit is regular, then $\he^h, \, \he^v$ lies in orthogonal subspaces of $\mf{g}$ in the sense of \eqref{eq:orth_decomp}, and it is always possible to find a  metric of the form in \eqref{eq:mertic} on $N\mc{O}$. If the orbit is degenerate, the situation is more subtle.
 If the metric still is block diagonal, the only situation where \eqref{eq:cross_terms} is not trivially satisfied is when both $\sigma_a,\sigma_b\in \ann(x)$, in which case it takes the form:
   \begin{equation*}
 g([n, \sigma_a], \sigma_b)+ g(\sigma_a, [n, \sigma_b]) 
 ,\end{equation*}
 which vanishes \emph{if} $g$ restricted to $\ann(x)$ is $\Ad_{H}$-invariant, since $n\in \ann(x)$. It is however not a priori true that there is such an $\Ad_H$-invariant metric, satisfying \eqref{eq:mertic}. In the case when $H$ is abelian, i.e. the orbit is regular, this can be shown to exist by straight-forward calculations. If $H$ is non-abelian, the situation is more subtle and the existence of such a metric needs to be verified.
 
It should be pointed out that the vectors $\he^\pm$ in general are not killing vectors of the metric in \eqref{eq:mertic}. However, the $\he^v$ are, which can be seen by a straight-forward calculation using the commutation relations of \eqref{eq:Lie_rel}:
   \begin{align}
0 =& \mc{L}_{\he^v_c} \left(  g (\he^+_a, \he^+_b) \right)   = 
\cancel{f_{cab}}
+
\cancel{f_{cba}}
+  i_{\he^+_a}i_{\he^+_b} \mc{L}_{e^v_c}  g 
\qquad \Rightarrow \quad  \mc{L}_{e^v_c}  g  = 0
 .\end{align}

The two sets of frames, $\he^\pm$ are related by an orthogonal matrix $P_{a}\,^b$  according to $\he^+_a = P_{a}\,^b \, \he^-_b$. This matrix encodes the inner product between the elements of the different frames, and is symmetric whenever $\mc{O}$ is regular.


\subsection{Some basic $O(d,d)$ generalised geometry}
In order to proceed, we need to introduce some basic concepts of generalised geometry. On a manifold $M$, we denote the generalised tangent bundle by  $E\sim TM\oplus T^*M$. Elements of $E$ are generalised vectors,  $V$,$W$, given by
\begin{align*}
V=
\begin{pmatrix}
v 
\\
\lambda
\end{pmatrix}
\qquad,\qquad
W=
\begin{pmatrix}
w 
\\
\mu
\end{pmatrix}
,\end{align*}
where $v,w \in TM$ denote ordinary vectors and $\lambda,\mu\in T^*M$ are one-forms.
On this space, we define the generalised Lie derivative as:
\begin{align*}
L_{V} W := 
[v,w]+\mathcal{L}_v \mu -i_{w}(\dd \lambda)
= 
[v,w]
+
\mathcal{L}_v \mu
 - \mathcal{L}_w \lambda  + \dd (  i_w  \lambda)
,\end{align*}
and we have a natural inner product (given by the $O(d,d)$-metric):
\begin{align}
\label{eq:nat_inner_prod}
\eta(V,W):=\frac{1}{2}\left( i_{v}\mu+i_w\lambda \right)
,\end{align}
where $i_{v}\mu$ denotes contraction on the first index.

In addition, there is also a generalised metric, $\mc{G}$, which, in addition to the metric $g$ depends on a two-form, $B$, and can be written in matrix notation as:
\begin{align}
\label{eq:gen_metric}
2 \mc{G}\; =&\;
\begin{pmatrix}
g - Bg^{-1}B & -Bg^{-1}
\\
g^{-1}B & g^{-1} 
\end{pmatrix}
\; = \;
(e^{-B})^T \mc{G}_0 e^{-B} 
\; := \; 
\begin{pmatrix}
 \mathds{1} & -B 
\\
0 &  \mathds{1}
\end{pmatrix}
\begin{pmatrix}
g & 0
\\
0 & g^{-1}
\end{pmatrix}
\begin{pmatrix}
 \mathds{1} & 0
\\
B & \mathds{1}
\end{pmatrix}
.\end{align}
This two-form $B$ determines the map from  $TM\oplus T^*M$  to $E $ via the exponential map:
\begin{align*}
e^B V:= v+\lambda+i_{v}B
.\end{align*}

A convenient formula for our purposes will be the relation between the generalised Lie derivative and the exponential map:
\beq
\label{eq:Liebniz_B}
L_{e^B V} e^B W = e^B \left(L_V W \right) -i_v i_w \left( \dd B\right)
.\eeq

\section{The generalised frame}
 
We can define a  generalised frame consisting of the ordinary frames $\he^{\pm}$ by:
 \begin{align}
\label{eq:frame}
\hat{E}_A
\; := \;
 \begin{pmatrix}
\hat{E}_a^+
\\
\hat{E}_a^-
\end{pmatrix}
\; = \; 
e^B \begin{pmatrix}
\he^+_a + i_{\he^+_a} g
\\
\he^-_a -  i_{\he^-_a} g
\end{pmatrix}
,\end{align}
where $i_{\he^\pm_a} g$ are the dual one-forms of the vectors $\he^\pm_a$.

It is straight-forward to show that these generalised frames are generalised orthonormal, i.e. they satisfy: 
{\small
\begin{align*}
 \mc{G}(\hat{E}_a^\pm, \hat{E}_b^\pm) =& \frac{1}{2 } \left(   g(\he^\pm_a, \he^\pm_b) +  g(\he^\pm_a, \he^\pm_b)  \right) = \delta_{ab}
 \\
  \mc{G}(\hat{E}_a^\pm, \hat{E}_b^\mp) =& \frac{1}{2 } \left(   g(\he^\pm_a, \he^\mp_b) -  g(\he^\pm_a, \he^\mp_b)  \right) = 0
  \\
 \eta(\hat{E}_a^\pm, \hat{E}_b^\pm) =& \frac{1}{2 } \left(   \pm g(\he^\pm_a, \he^\pm_b)  \pm  g(\he^\pm_b, \he^\pm_a)  \right) = \pm \delta_{ab}
 \\
 \eta(\hat{E}_a^\pm, \hat{E}_b^\mp) =& \frac{1}{2 } \left(   \mp  g(\he^\pm_a, \he^\mp_b) \pm  g(\he^\pm_a, \he^\mp_b)  \right) = 0
\end{align*}
}

However, for the purposes of consistent truncations, what we really need is to find a \emph{generalised Liebniz frame}, so we want our generalised frame to satisfy the commutation relations of equation \eqref{eq:comm_rels_pm}, that is:
{\small
\begin{align}
\label{eq:Liebniz_gen}
L_{\hat{E}^\pm_a} \hat{E}^\pm_b = & 
\frac{1}{2} f_{ab}\,^c \left( 
3\hat{E}^\pm_c
-
\hat{E}^\mp_c
\right)
\quad ,\quad
L_{\hat{E}^+_a} \hat{E}^-_b = 
\frac{1}{2} f_{ab}\,^c  \, \left( 
\hat{E}^+_c
+
\hat{E}^-_c
\right)
.\end{align}
}

 The vector parts of these relations are satisfied automatically, but not the one-form parts. We therefore define a three-form  by assuming $H=\dd B$ together with equation \eqref{eq:Liebniz_B}. That is, we define $H$ by requiring the Liebniz relations \eqref{eq:Liebniz_gen} to hold. One can show that these all give equivalent definitions for $H$ using the Jacobi identity, so it is enough for us to consider one, say   for example $L_{\hat{E}^+_a} \hat{E}^+_b$:
 {\footnotesize
 \begin{align}
 L_{\hat{E}^+_a} \hat{E}^+_b  =& \frac{1}{2} f_{ab}\,^c \left( 
3\he^+_c
-
\he^-_c
\right)
+\mc{L}_{\he^+_a} i_{\he^+_b}g -i_{\he^+_b}  d (  i_{\he^+_a} g)   -i_{\he^+_a}i_{\he^+_b}H 
\,
\overset{!}{ =}  \, 
 \frac{1}{2} f_{ab}\,^c \left[ 
3 \left( 
\he^+_c+  i_{\he^+_c} g
\right)
-
\left( 
\he^-_c -  i_{\he^-_c} g
\right)
\right]
 \end{align}
 }
 so
 \begin{align}
  i_{\he^+_a}i_{\he^+_b}H 
  =
\mc{L}_{\he^+_a} i_{\he^+_b}g -i_{\he^+_b}  d(  i_{\he^+_a} g )
-
\frac{1}{2} f_{ab}\,^c \left( 
3 i_{\he^+_c} g
+ i_{\he^-_c} g
\right)
. \end{align}
If this 3-form $H$ is closed, this indeed gives us a two-form $B$ via $\dd B = H$, and the frame $\hat{E}^\pm$ defines a generalised Liebniz prarallellisation of our space $N \mc{O}$.

 Since $\he^+$ is a globally defined frame, we can just as well define $H$ through:
  \begin{align}
i_{\he^+_c} i_{\he^+_b}   i_{\he^+_a}H 
  =
i_{\he^+_c} i_{\he^+_b}  d(  i_{\he^+_a} g)
+ 
i_{\he^+_c} \frac{1}{2} f_{ab}\,^d \left( 
3 i_{\he^+_d} g
+ i_{\he^-_d} g
\right)
- i_{\he^+_c}  \mc{L}_{\he^+_a} i_{\he^+_b}g 
. \end{align}
 To compute this is relatively straight forward,  and results in:
{ \small
   \begin{align*}
i_{\he^+_c} i_{\he^+_b}   i_{\he^+_a}H 
=&
- \frac{3}{2} f_{abc}
+\frac{1}{2}  f_{bc}\,^d P_{da}
+ \frac{1}{2} f_{ab}\,^d P_{dc}
+  \frac{1}{2} f_{ca}\, ^d P_{bd}
. \end{align*}
 }
 
So the three-form $H$ is given by
 {\footnotesize
 \begin{align}
 \label{eq:H}
 H =\frac{1}{12} \left[ 
- 3f_{abc}
+  f_{bc}\,^d P_{ad}
+  f_{ca}\, ^d P_{bd}
+  f_{ab}\,^d P_{cd}
\right] 
\;
i_{\he^+_a} g  \wedge  i_{\he^+_b}g  \wedge  i_{\he^+_c}g
. \end{align}
} 
 Using the Jacobi identity, both in $\mathfrak{g}$ and in $\text{Lie}(G\ltimes \mf{g})$, one can show that this is indeed a closed form, and so this defines a good two-form $B$ via $H=dB$ (and it indeed satisfies the other Liebniz relations as well).  
This calculation is spelled out in appendix \ref{app:dH=0} and  hinges vitally on the orthogonality of the matrix $P$ together with that  the metric indeed can be defined uniquely by \eqref{eq:mertic}.

\section{Some examples \label{sec:ex}}
It is well-known that the normal bundle to a regular adjoint orbit is globally flat \cite{terng1985}. Below, we will consider some examples, including both regular and degenerate orbits. Such orbits have been studied in for example \cite{rudolph2017differential} for some low-dimensional examples,  and  in \cite{bernatska2008} for $G=SU(n),SO(n)$ and $Sp(n)$.

\subsection{$G=SU(2)$  \label{sec:ex_SU(2)}}
In this case, there is only one type of adjoint orbit which, as a manifold, is isomorphic to $S^2$. This is a regular adjoint orbit and its normal bundle hence trivial, allowing us to write the space $N\mc{O}^{SU(2)}$ as a global product:
\beq
N\mc{O}^{SU(2)}=S^2\times\mbb{R}
.\eeq
 This space was first shown to admit a consistent truncation of the bosonic string in \cite{Cvetic:2003jy}, and  an explicit generalised $O(d,d)$-frame was constructed on this space in\cite{deFelice:2014}, satisfying the Leibniz conditions as required.
 To make contact with the expressions presented therein, we can express the three-form $dB$ in terms of the vectors $\he^{h,v}$ as:
   {\footnotesize
 \begin{align*}
 dB =& \frac{1}{2^5 \cdot 3} \left[- 3f_{abc}
+  f_{bc}\,^d P_{ad}
+  f_{ca}\, ^d P_{bd}
+  f_{ab}\,^d P_{cd}
\right] 
\\
\;  &
 i_{\he^h_a} g
\wedge 
 i_{\he^h_b} g
\wedge 
 i_{\he^h_c} g
+
3\, 
 i_{\he^h_a} g
\wedge 
 i_{\he^h_b} g
\wedge 
 i_{\he^v_c} g  
+
3\, 
 i_{\he^v_a} g  
\wedge 
  i_{\he^v_b} g 
\wedge 
i_{\he^h_c} g
+
 i_{\he^v_a} g  
\wedge 
  i_{\he^v_b} g 
\wedge 
  i_{\he^v_c} g 
. \end{align*}
}
Here,  $\mf{h} =\mf{u}(1)$ is one-dimensional, so three-form becomes:
    {\footnotesize
 \begin{align*}
 dB_{SU(2)} =& \frac{1}{2^5 \cdot 3} \left[- 3f_{abc}
+  f_{bc}\,^d P_{ad}
+  f_{ca}\, ^d P_{bd}
+  f_{ab}\,^d P_{cd}
\right] 
\\
\;  & \qquad
3\, 
 i_{\he^v_a} g  
\wedge 
  i_{\he^v_b} g 
\wedge 
i_{\he^h_c} g
\CGR{+
 i_{\he^v_a} g  
\wedge 
  i_{\he^v_b} g 
\wedge 
  i_{\he^v_c} g 
  }
. \end{align*}
}
 Furthermore, since the vectors $\he_a^v$ are given in terms of structure constants, (which is clear from the commutator in equation \eqref{eq:vecs}), and since these in the $SU(2)$ case are simply Levi-Civita's, the product of these in $dH$ may be expanded in terms of delta-functions,  after which it is clear that the term \CGR{$ i_{\he^v_a} g   \wedge    i_{\he^v_b} g  \wedge   i_{\he^v_c} g $}-term above vanish, and the only surviving part of $ dB_{SU(2)} $ is the mixed term between $\he^h$ and $\he^v$, which indeed reproduces the expressions in \cite{deFelice:2014}

  The $\mbb{R}$-factor can  be compactified to an $S^1$, allowing for a straight-forward compactification to be carried out on such an internal space, giving rise to a consistent truncation.

\subsection{$G=SU(3)$}
Here, there are two classes of orbits, the regular orbit $\mc{O}_{reg}^{SU(3)} \, = \, \frac{SU(3)}{U(1)\times U(1)} \simeq \mbb{F}^3$,  which is the six-dimensional flag manifold $\mbb{F}^3$ \cite{rudolph2017differential} 
, and the degenerate orbit $\mc{O}_{deg}^{SU(3)} \,= \, \frac{SU(3)}{SU(2)\times U(1)} \simeq \CP^2$.
For the regular orbit, the stabiliser subgroup is abelian, and our construction of the generalised frame is straight-forward, arriving at a generalised Liebniz frame on: 
\beq
N\mc{O}_{reg}^{SU(3)}=\mbb{F}^3 \times \text{T}^2
.\eeq

The normal bundle over the degenerate orbit, $N\mc{O}_{deg}^{SU(3)}$, will no longer be a global product, but rather consist of an  $SU(2)\times U(1)$ principal bundle over a $\CP^2$ base, giving us:
\beq
N\mc{O}_{deg}^{SU(3)}=\pi:SU(3)\rightarrow \CP^2 
.\eeq

\subsection{$G=SO(4)$}
This is not a very interesting situation since it's double cover is given  by $SU(2)\times SU(2)$. Therefore, the regular adjoint orbits of $SO(4)$ will be diffeomorphic to $S^2\times S^2$, so this will give us:
\beq
N\mc{O}_{reg}^{SO(4)} = S^2\times S^2\times \text{T}^2
.\eeq
The only non-trivial degenerate orbit will be diffeomorphic to $S^{2}$, and the manifold $N\mc{O}^{SO(4)}_{deg}$ will then consist of  $SU(2)\times U(1)$-fibres over this.

\subsection{$G=SO(5)$}
Here, the regular orbit is given by
\beq
N\mc{O}_{reg}^{SO(5)}  = \frac{SO(5)}{SO(2)\times SO(2)} \times \text{T}^2
\eeq
where the space $\frac{SO(5)}{SO(2)\times SO(2)} $ is a real, 'flag-like manifold'\cite{Boya:2001}. There are a degenerate orbit corresponding to the real grassmannian $\mbb{R}G_{3,2}$, whose normal space will be a principal $SO(3)\times SO(2)$-bundle over $\mbb{R}G_{3,2}$.  
 
It could be argued that since  $\dim(SO(5)) =10$, that this is not a very interesting example from the point of view of consistent truncations. However,  $O(d,d)$ generalised geometry captures the bosonic degrees of freedom, and should therefore also apply to bosonic string theory, making higher dimensions interesting.

\section{Discussion and Outlook }
The novel point of this note was to give an explicit construction of a generalised Liebniz frame of a new class of spaces given by normal bundles over adjoint orbits $\mc{O}$ of some group $G$.
 In the cases where the normal bundle is flat, this construction is complete and the normal directions can be compactified, and these manifolds should admit consistent truncations according to the recipe presented in \cite{Lee:2014mla}. In the case when this bundle is not trivial, the construction hinges on the existence of a suitable metric according to \eqref{eq:mertic}, and the compactification procedure required to obtain a low-energy effective theory is, at best, very subtle and is beyond the scope of this note.

 Another line of attack closely related to the one presented herein is to think of the space $\mc{O}\sim G/H$ as  a  submanifold   of  a local group manifold $M\sim G$, and ''zooming in'' very close to such an orbit. If one zooms in enough, the manifold should look like $N\mc{O}$. Since it is possible to construct a generalised frame on a local group manifold, it is reasonable to imagine this should also hold true on $N\mc{O}$ by such a construction, and the moduli one obtain in this way should be a limit of the moduli on the group manifold $G$. In the case when the stabiliser is abelian, it is  indeed possible to carry out a similar construction to the one herein by considering In\"on\"u-Wigner contractions\cite{Inonu:1953} of the group $G$ (and the stabiliser $H$), together with a rescaling of the metric and the normal coordinates. This procedure, however, appears to be dependent on a global splitting of the metric along, and orthogonal to, the orbit, which can only be done in the case where $H$ is abelian.

In the recent work of \cite{Inverso:2017lrz},   the scope of possible generalised Scherk-Schwarz reductions is elegantly considerd. This approach differs in that it is not primarily focused on consistent truncations but rather generalised Leibniz parallelisability on its own. The manifolds arising from regular orbits in this note should fall within the classification presented in \cite{Inverso:2017lrz}, though the degenerate ones does not. It would be interesting to see if any connection can be made with the manifolds arising as normal bundles of degenerate orbits as well, and how the issues that arise in this construction is reflected in the setting of \cite{Inverso:2017lrz}. This is left to future work.

\section*{Acknowledgments}
I am very grateful for the many helpful discussions with Daniel Waldram, and the initial collaboration with Raphael Cavallari. This work was supported by the  EPSRC programme grant ``New Geometric
Structures from String Theory'', EP/K034456/1.

\appendix
\section{Showing the three-form defined in \eqref{eq:H} is closed \label{app:dH=0}}
We wish to show that the 3-form 
 {\footnotesize
 \begin{align*}
 H = \frac{1}{12} \left[ 
-3f_{abc}
+  f_{bc}\,^d P_{da}
+  f_{ca}\, ^d P_{db}
+  f_{ab}\,^d P_{dc}
\right] 
\;
i_{\he^+_a} g  \wedge  i_{\he^+_b}g  \wedge  i_{\he^+_c}g
, \end{align*}
} 
is closed. 
 To compute $dH$, consider first 
 \begin{align*}
 i_{\he^+_c} i_{\he^+_b}d( i_{\he^+_a}g) = &
-
  i_{\he^+_a} g(  [ \he^+_b , \he^+_c ])
    =
  -\frac{1 }{2}f_{bc}\,^d  \left( \delta_{da}- P_{da}\right)
 . \end{align*}
%
This gives us:
  {\scriptsize
 \begin{align*}
 dH
= &
 -\frac{3}{2}f_{mn}\,^e \left( 3\delta_{ea} -P_{ea} \right)
 \Big(
-3f_{abc} 
+  f_{bc}\,^d P_{da}
+  f_{ca}\, ^d P_{db}
+   f_{ab}\,^d P_{dc}
\Big)
 \;\;
  i_{\he^+_m}g\wedge  i_{\he^+_n}g \wedge  i_{\he^+_b}g \wedge i_{\he^+_c} g  
\;
\\ = &
 \Big[
 -\frac{3^2}{2}f_{mna}\,
 \Big(
 - \xcancel{3f_{abc} }
+  f_{bcd} P^{ad}
+ \xcancel{ f_{ca}\, ^d P_{db}}
+ \xcancel{   f_{ab}\,^d P_{dc}}
\Big)
\\ & \quad
 +\frac{3}{2}f_{mn}\,^e P_{ea} 
 \Big(
 -3f_{abc} 
+  f_{bc}\,^d P_{da}
+  f_{ca}\, ^d P_{db}
+   f_{ab}\,^d P_{dc}
\Big)
\Big]
 \;\;
  i_{\he^+_m}g\wedge  i_{\he^+_n}g \wedge  i_{\he^+_b}g \wedge i_{\he^+_c} g  
  \\
  =&
\frac{3}{2} \Big[
 -3f_{mna}  P^{ad} f_{bcd} 
 -3 f_{mnd} P^{da} f_{abc} 
+\CGR{  f_{mn}\,^e f_{bcd}\,^d  P_{ea}  P_{da}  }
+
f_{mnd}P^{da}
\left(  \,  f_{ca}\, ^e P_{eb}
+\,  f_{ab}\,^e P_{ec}
\right)
\Big]
 \;\;
  i_{\he^+_m}g\wedge  i_{\he^+_n}g \wedge  i_{\he^+_b}g \wedge i_{\he^+_c} g  
  \\
 \end{align*}
}
 where \CGR{the grey term} vanish due to the  Jacobi identity in $G$ since $P$ is orthogonal.
 The other terms  add up, giving us in total:
  {\scriptsize
 \begin{align*}
 dH =&
-3 f_{mnd}  P^{da}
 f_{ab}\, ^e
  \Big[
 3 \delta_{ec}
-
 \,  P_{ec}
\Big]
 \;\;
  i_{\he^+_m}g\wedge  i_{\he^+_n}g \wedge  i_{\he^+_b}g \wedge i_{\he^+_c} g  
\end{align*}
}

But, from requiring the Jacobi identity to hold on $\text{Lie}(G\ltimes \mf{g})$, that is
{\footnotesize
\beq
0=
 [\he^+_a, [\he^+_b, \he^+_c ]]
  +
 [\he^+_b, [\he^+_c, \he^+_a ]]
   +
 [\he^+_c, [\he^+_a, \he^+_b ]] 
,\eeq
}
we find a Jacobi identity that states:
{\footnotesize
 \begin{align*}
0=&
 P^{da} 
 \Big(
f_{mnd} f_{cae}
  +
f_{ncd}  f_{mae}
   +
f_{cmd}    f_{nae}
 \Big) 
 \Big[
3 \delta^{eb} -P\,^{eb}
 \Big]
 \\
 &\Longrightarrow
 \\ &
f_{mnd}   P^{da}  f_{cae}
\Big[
3 \delta^{eb} -P\,^{eb}
 \Big]
 =  -P^{da} 
 \Big(
f_{ncd}  f_{mae}
   +
f_{cmd}    f_{nae}
 \Big) 
 \Big[
3 \delta^{eb} -P\,^{eb}
 \Big]
, \end{align*}
 }
 which gives us 
  \begin{align}
 dH = -2dH 
 \quad
\Rightarrow
\quad dH=0
\end{align}


{\scriptsize
\bibliographystyle{JHEP}
\bibliography{references.bib}
}

\end{document}